\documentclass[twocolumn,showpacs,aps,prl,floatfix]{revtex4}

\usepackage{graphicx}
\usepackage{epsfig}
\usepackage{graphics,color}
\usepackage{amsmath}
\usepackage{amsfonts}
\usepackage{amssymb}
\usepackage{subfigure}

\newcommand{\bra}[1]{\langle #1|}
\newcommand{\ket}[1]{|#1\rangle}

\newcommand{\CC}{{\cal C}}

\begin{document}

\title{Unified description of charge transfer mechanisms and \\
vibronic dynamics in nanoscale junctions}
\author{R. Avriller${}^{1,2}$}
\affiliation{${}^{1}$Donostia International Physics Center (DIPC),
E-20018 Donostia-San Sebasti\'{a}n, Spain. \\ 
${}^{2}$Departamento de F\'{\i}sica Te\'orica de la Materia
Condensada C-XII, Facultad de Ciencias, Universidad Aut\'onoma de Madrid,
E-28049 Madrid, Spain.}
\date{\today}
\begin{abstract}
We propose a general framework that unifies the point of view of counting
statistics of transmitted (fermionic) charges as it is commonly used in the quantum transport
community to the point of view of counting statics of phonons (bosons).
As a particular example, we study on the same footing the counting
statistics of electrons transferred through a molecular junction and the
corresponding population dynamics of the associated molecular vibrational
mode. In the tunnel limit, non-perturbative results in the electron-phonon
interaction are derived that unify complementary approaches based on rate
equations or on the use of non-equilibrium Green functions.  
\end{abstract}


\maketitle

\section{I. Introduction}

The concept of counting statistics of transmitted charges as it was first
defined and computed for non-interacting tunnel junctions by Levitov
\textit{et al} \cite{Hamiltonian_FCS_Levitov_Formula,Hamiltonian_FCS_Levitov_Formula_2} is based on an analogy
between coherent electronic transport and quantum
optics \cite{Photon_Correlations}. In both fields of research, the
underlying electronic or photonic quantum field undergoes quantum fluctuations
encoded in a fundamental quantity, namely the probability distribution
$P(q)$ that \textit{$q$ charges are transferred through the junction}
or the stationary probability distribution $P(n)$ of \textit{detecting $n$ photons in the
electromagnetic field}, in a given amount of time. \\
This concept has proved to be a powerful one
\cite{Quantum_Noise_Mesoscopic_Physics} by providing deep connections between the 
fields of molecular electronics
\cite{Electron_Transport_Molecular_Wire_Junctions} and quantum noise
\cite{Shot_Noise_Mesoscopic_Conductors}. On a more fundamental point of view,
the concept of counting statistics of transmitted charges drew the attention
of the condensed-matter community to the idea already present in quantum optics,
that the quantum nature of transport mechanisms as well as the peculiar
effects related to quantum mechanics are
more generally encoded into correlation functions of the fields rather than
into the mean value of a given observable. \\
Recently, ideas of measuring phonon shot noise have been reported in the
literature \cite{Quantum_Measurement_Phonon_Shot_Noise}, but interestingly,
there is to our knowledge no work that
considers both kinds of descriptions on the same footing. 
We thus assign the following goals to the present article. It is first to provide such a
general framework, that unifies the point of view of
counting statistics of transmitted (fermionic) charges as it
is currently accepted in the field of quantum transport, to
the one of counting statistics of phonons (bosons) inherited from the quantum
optics community. Then,
it is to illustrate the fertility of such a framework by studying the concrete example
of a quantum transport problem through an interacting nanoscale device. The
corresponding issue is indeed relevant to the quantum transport community,
where we can distinguish between two main approaches to derive the associated transport
properties. \\
On one side, the study of single electron transistors
has motivated theoretical investigations of sequential tunnelling of charges through local
devices, in presence of interactions with the electromagnetic environment
\cite{Charge_Tunneling_Nanojunctions}  or with internal degrees of freedom
(electron-electron or electron-phonon interactions). The corresponding
theoretical framework based on rate equations with transition rates derived
from the Fermi golden rule has been successfully applied in more recent works
related to the problem of Franck-Condon blockade \cite{FC_Blockade_Regime} and
observed experimentally in suspended carbon nanotubes
\cite{FC_Blocakde_Suspended_CNTs}. In presence of electron-phonon (e-ph)
interactions, the obtained current versus voltage curves
\cite{Transport_I_V_molecules} exhibit characteristic 
patterns (satellite peaks at voltages multiple of twice the phonon energy) due
to the activation of tunnelling assisted by phonon emission, as well as a
reduction of the low-bias conductance (Franck-Condon blockade). The underlying
transport mechanism was shown to induce giant Fano factors for the
fluctuations of the transmitted charges (current noise), when the vibrational
population was driven far into the non equilibrium regime
\cite{FC_Blockade_Giant_Fano_Factors}. The corresponding counting statistics
of transmitted charges was derived for molecular junctions in Ref. \cite{FCS_Low_Gamma,FCS_Low_Gamma_2} 
and for nanoelectromechanical systems in Ref. \cite{FCS_charge_shuttle}. \\
On the other side, the experimental works related to coherent transport
through atomic chains \cite{Atomic_Chains} and molecular junctions
\cite{Molecular_junctions} has driven a tremendous amount of theoretical work,
based on the extensive use of Keldysh non-equilibrium-Green-functions
formalism \cite{Keldysh_Formalism}. In both approaches based on model
Hamiltonians
\cite{Quantal_Transistor,Inelastic_electron_tunneling_spectroscopy,Voltage_induced_singularities_transport}
or on more sophisticated \textit{ab-initio} calculations
\cite{Inelastic_Scattering_Local_Heating_Atomic_Gold_Wires,Electron-vibration_interaction_transport_atomic_gold_wires,Universal_features_electron_phonon_interactions_atomic_wires}, the conductance
characteristics exhibit a jump at the inelastic threshold (voltages
corresponding to the phonon energy) associated to the activation of
tunnelling with emission of a phonon, the sign of which is determined by the
transmission coefficient of the junction. Such feature is of great experimental
interest in order to perform an inelastic spectroscopy of the device
\cite{Inelastic_electron_tunneling_spectroscopy} (spectroscopy of the
vibrational modes, measurement of the e-ph coupling strength from the height of the
jump of conductance). More recently, some
theoretical works focused on inelastic signatures of e-ph interactions on noise
characteristics \cite{Theory_current_shot-noise,Inelastic_tunneling_effects_noise} and on the full counting statistics of the transmitted
charges in the coherent regime \cite{FCS_Molecular_junctions,FCS_Molecular_junctions_2,FCS_Molecular_junctions_3}. The appearance of jumps in the
derivative of the current noise versus voltage curves was shown to result from a competition
mechanism between elastic and inelastic processes of tunnelling, and the feedback of the phonon dynamics 
to be an important effect in order to correctly capture the behaviour of higher order cumulants in the regime 
of large voltages \cite{FCS_Molecular_junctions_Role_Phonon_Fluctuations}. \\

The organisation of this article is then the following.
In the first part, we provide a general framework that unifies the
point of view of counting statistics of transmitted (fermionic) charges in nanodevices to
the counting statistics of phonons (bosons). As a particular case of our formalism, we then consider in
detail the particular model of a nanoscale junction as a single molecular
level coupled to a local vibrational (phonon) mode and to perfect
reservoirs with different chemical potentials (non equilibrium situation). We
show that both approaches based on rate equations in the sequential tunnelling
regime or on non-equilibrium-Green-functions in the coherent regime may be
described in a unified way following the scope of our formalism. We interpret
the transport properties in such a device as the result of a
self-consistent mechanism between charge transfer through the junction and the
dynamics of population of the local phonon mode. As an illustration, the full counting statistics (FCS) of
transferred electrons and phonons are defined and computed exactly in
the tunnel limit.

\section{II. General framework}

\subsection{A. The model}

\begin{figure}[!ht]
   \begin{center}
        \includegraphics[width=0.85\columnwidth]{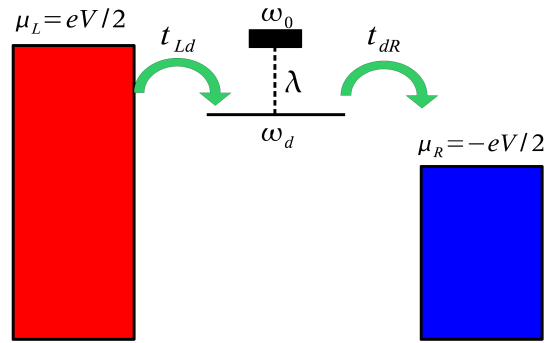}
        \caption{Representation of the nanoscale junction.
        \label{fig1:Diagrams}}
 \end{center}
\end{figure}

In the following, we consider the simple model of a molecular junction (see
Fig.\ref{fig1:Diagrams}) described by the Hamiltonian (in units
$e=\hbar=k=1$)

\begin{eqnarray}
H &=& H_{mol} + \sum_{X=L,R} H_{X} + H_{T} 
\label{eqn:Eq1} 
\end{eqnarray}

\noindent
In Eq.\ref{eqn:Eq1}, the molecular system is encoded into the Hamiltonian
$H_{mol} = (\omega_d + \lambda \phi)n_d + \omega_0 a^{\dagger}a$ that describes a
single molecular (dot) level of energy $\omega_d$ interacting with a local
phonon (vibrational) mode of energy $\omega_0$. The electron-phonon
interaction is proportional to the electronic density operator
$n_d=d^{\dagger}d$ and to the vibronic position operator $\phi = a +
a^{\dagger}$. It is characterised by the e-ph coupling strength $\lambda$. 
The second term in Eq.\ref{eqn:Eq1} is the Hamiltonian of non interacting left (right)
leads $H_{L(R)}$ both maintained under a symmetric potential drop
$\mu_L=-\mu_R=V/2$.
In such a system, charge might be transferred from the electrodes to the molecule by the
tunnel Hamiltonian $H_{T} = (t_{Ld}\Psi_L^{\dagger} + t_{Rd}\Psi_R^{\dagger})d
+ \mbox{hc}$ that couples the dot level to the $X=L,R$ lead
through the hopping term $t_{Xd}$. The typical energy scale corresponding to this charge
transfer mechanism is given by the $X=L,R$ tunnelling rate
$\Gamma_{X}=|t_{Xd}|^2/W$, where in the wide band approximation, the bandwidth $W=1/\pi\rho_0$ is
inversely proportional to the flat density of states of the leads $\rho_0$. 
\\
Although simple, the model described by Eq.\ref{eqn:Eq1} exhibits a non
trivial phase diagram characterised by two competing energy scales \cite{Quantal_Transistor}, namely
$\Gamma=\Gamma_L+\Gamma_R$ the total coupling strength to the leads
and $\lambda^2/\omega_0$ the e-ph coupling strength.
In this article, we concentrate on the regime of weak tunnelling for which
$\Gamma \ll \lambda^2/\omega_0$ is the smallest energy scale of the
problem. In this regime, the e-ph interaction strongly normalises
electronic degrees of freedom and a non-perturbative approach in the e-ph
coupling strength is needed. \\

We further perform a unitary Lang-Firsov (polaron) transformation \cite{Lang_Firsov_Transformation}
$U=e^{gn_d(a-a^{\dagger})}$ in order to explicitly eliminate the e-ph interaction
term from $H_{mol}$ by shifting the vibronic position operator $\phi$
($g=\lambda/\omega_0$ is the dimensionless e-ph
coupling strength). The obtained dual representation is then more adapted
for perturbation calculations in leading orders of $\Gamma$, and the
transformed Hamiltonian $\tilde{H}=UHU^{\dagger}$ now reads 
 
\begin{eqnarray}
\tilde{H} &=& \tilde{H}_{mol} + \sum_{X=L,R} H_{X} + \tilde{H}_{T} 
\label{eqn:Eq2}  
\end{eqnarray}

\noindent
In Eq.\ref{eqn:Eq2}, the Hamiltonian describing the molecular system 
$\tilde{H}_{mol} = \tilde{\omega}_d n_d + \omega_0 a^{\dagger}a$ does not
couple anymore electronic and vibrational degrees of freedom, and the dot
position is renormalised by the polaronic shift $\tilde{\omega}_d=\omega_d -
g^2\omega_0$. Unfortunately, the price to pay in the Lang-Firsov
transformation is that the transformed tunnelling operator $\tilde{H}_{T} =
\lbrace t_{Ld}\Psi_L^{\dagger} + t_{Rd}\Psi_R^{\dagger} \rbrace d X +
\mbox{hc}$ acquires a phase operator $X=e^{g(a-a^{\dagger})}$ that controls
the injection of charges in the system.

\subsection{B. Counting statistics of electrons and phonons}

The charge transfer mechanism in the molecular junction shown in
Fig.\ref{fig1:Diagrams} is a coherent, time
dependent process involving tunnelling of charges from the electrodes to the
dot and energy exchange with the local phonon mode (emission and absorption of
phonons). In the steady state, the electronic current (rate of charge
transfer per unit of time) and the vibrational population are 
constant in average. However, both quantities might fluctuate in time. The most complete
information about those fluctuations is encoded in the joined probability
distribution $P(q,n)$, defined as the probability that \textit{$q$ charges are
transferred through the junction} and \textit{the number of phonons populating the mode has
varied by an amount of $n$ quanta} during the measuring time $t_0$ (this time is large enough to reach the stationary
state). This quantity generalises the notion of full counting statistics (FCS)
of transmitted charges which is recovered by tracing out the vibrational
degrees of freedom $P_{el}(q) = \sum_{n\in\mathbb{Z}} P(q,n)$. Similarly, one
could obtain a phonon counting statistics by tracing out the electronic
degrees of freedom $T_{ph}(n) = \sum_{q\in\mathbb{Z}} P(q,n)$. In general, the
joined distribution cannot be factorised due to the presence of e-ph
correlations, \textit{i.e.} $P(q,n) \neq P_{el}(q) T_{ph}(n)$.\\  

The aim of this article is to provide a way of computing this
joined distribution by treating electronic and vibronic fluctuations on the
same footing. By analogy with the electronic case \cite{Hamiltonian_FCS_Levitov_Formula,Hamiltonian_FCS_Levitov_Formula_2,FCS_Gogolin}, we define the cumulant
generating function (CGF) as the Fourier transform of the distribution
$P(q,n)$, namely $S(\chi,\xi)=-\ln\lbrace
\sum_{q\in\mathbb{Z}}\sum_{n\in\mathbb{Z}} e^{i(q\chi + n\xi)}P(q,n)
\rbrace$. This functional generates the cumulants of the distribution $P(q,n)$ by successive
derivations of the CGF with respect to the electronic (vibrational) counting
field $\chi$ ($\xi$). The connection of this quantity to the Hamiltonian of
Eq.\ref{eqn:Eq2} is made by computing the following contour-ordered evolution
operator in Keldysh space  

\begin{eqnarray}
S(\chi,\xi) = -\ln \Big{\langle} \mbox{T}_c \exp\big{\lbrace} -i\int_c dt
\tilde{H}_{T;\chi(t),\xi(t)} \big{\rbrace} \Big{\rangle}_0
\label{eqn:Eq3}
\end{eqnarray}

\noindent
In Eq.\ref{eqn:Eq3}, the tunnelling operator is obtained by performing the
following substitutions in the $\tilde{H}_{T}$ Hamiltonian, namely $t_{L(R)d}
\rightarrow t_{L(R)d} e^{\mp i\chi(t)/4}$ and $a\rightarrow a e^{-i\xi(t)/2}$.
The mean value is taken with respect to the unperturbed Hamiltonian
$H_0 = \tilde{H} - \tilde{H}_{T;\chi(t),\xi(t)}$, and 
the electronic counting field $\chi(t)$ is equal to
$+(-)\chi$ on the positive (negative) branch of the Keldysh
contour $\CC$. It stands for a non destructive measurement of the
charge transferred from any electrode to the dot \footnote{In the
  low-transmission limit, the choice of gauge obtained by putting the
  electronic counting field into the left electrode only ($t_{Ld} \rightarrow
  t_{Ld} e^{-i\chi(t)/2}$) or equally shared into both electrodes ($t_{L(R)d}
  \rightarrow t_{L(R)d} e^{\mp i\chi(t)/4}$) are not equivalent. As explained
  in Appendix B, we chose the gauge preserving current conservation in the
  tunnel limit $\Gamma \rightarrow 0$.}. 
Similarly, the phonon counting field $\xi(t)$ is equal to $+(-)\xi$ on the positive (negative) branch of
the $\CC$-contour and stands for a virtual measurement of the energy transferred to the local
phonon mode, \textit{i.e.} the net variation of the number of vibrational
quanta during the charge transfer process.   

\subsection{C. Expansion of the CGF in the tunnel limit}

\begin{figure}[!ht]
   \begin{center}
        \includegraphics[width=0.60\columnwidth]{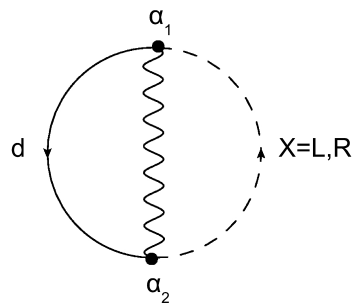}
        \caption{Linked cluster expansion of the CGF $\delta
          S^{(1)}(\chi,\xi)$ to lowest order in $\Gamma$. Plain (dashed) line
          corresponds to the dot (electrode) Green function. The wavy line
          stands for the phonon Green function. 
        \label{fig2:Bubble_Expansion}}
 \end{center}
\end{figure}

Evaluating Eq.\ref{eqn:Eq3} for any value of the parameters
$(\Gamma,\lambda^2/\omega_0)$ is a formidable task. As a particular case of
our general framework, we would like to address in detail the tunnel regime, for
which $\Gamma \ll \lambda^2/\omega_0, \omega_0$ and the charges are injected
sequentially into the dot. 
Our calculation holds for off-resonant (low-transmission) situations, for which $|\omega_d| \gg \Gamma$ and 
correlated transmission events (electron bunching) may be neglected. In this regime, 
the resulting electronic FCS is anticipated to be Poissonian, but as far as we
know, the explicit analytical evaluation of the FCS is still not reported in the
literature. \\

We thus perform a linked cluster expansion of the CGF to lowest order in the
tunnelling rate $\Gamma$, or in an equivalent manner to the lowest order in the
transmission coefficient (see Fig.\ref{fig2:Bubble_Expansion}). Even if restricted
to order $\Gamma$, this calculation holds for arbitrary values of the e-ph coupling strength $g$. The
corresponding expression of the CGF provides in a single, compact formula,
the most complete information about both electronic and vibrational
fluctuations in the tunnel limit

\begin{widetext}
\begin{eqnarray}
\delta S^{(1)}(\chi,\xi) = \frac{t_0}{2} W \sum_{\alpha_1\alpha_2=\pm} \alpha_1 \alpha_2 \int
\frac{d\omega_1}{2\pi} \int \frac{d\omega_2}{2\pi}
\mathcal{P}^{\alpha_1\alpha_2}_{\xi}(\omega_1) g^{\alpha_1\alpha_2}_{dd}(\omega_2)
\Big{\lbrace} \Gamma_{L;\chi}^{\alpha_2\alpha_1}
g^{\alpha_2\alpha_1}_{LL} + \Gamma_{R;\chi}^{\alpha_2\alpha_1}
g^{\alpha_2\alpha_1}_{RR} \Big{\rbrace}(\omega_1+\omega_2)
\label{eqn:Eq4}  
\end{eqnarray}
\end{widetext}   

\noindent
where the counting field dependent tunnelling matrix element $\Gamma_{L(R);\chi}^{\alpha_2\alpha_1} = \Gamma_{L(R)}
e^{\pm i(\alpha_2-\alpha_1)\chi/4}$ takes into account charge transfer
processes from the left (right)
electrode to the dot. Evaluation of Eq.\ref{eqn:Eq4}
involves the bare dot Green function $\hat{g}_{dd}(t) = -i \langle T_c d(t)
d^{\dagger}(0)\rangle_0$ (plain line in Fig.\ref{fig2:Bubble_Expansion}), the
$X=L,R$ lead Green function $\hat{g}_{XX}(t) = -i \langle T_c \Psi_X(t)
\Psi_X^{\dagger}(0)\rangle_0$ (dashed line in Fig.\ref{fig2:Bubble_Expansion})
and the $\xi$-dependent phonon Green function $\mathcal{P}_\xi(t)= \langle T_c X(t)
X^{\dagger}(0)\rangle_0$ (wavy line in
Fig.\ref{fig2:Bubble_Expansion}). Explicit expressions for those Green
functions are given in Appendix A.

\section{III. Results in the tunnel regime}

\subsection{A. Electronic FCS}

We derive from Eq.\ref{eqn:Eq4} an
analytical expression for the electronic CGF in the tunnel regime, namely $\delta
S^{(1)}_{el}(\chi) = \delta S^{(1)}(\chi,\xi=0)$

\begin{eqnarray}  
\delta S^{(1)}_{el}(\chi) = -t_0 \lbrace \Gamma_{L \rightarrow R}(e^{i\chi}-1) +
\Gamma_{R \rightarrow L}(e^{-i\chi}-1) \rbrace \label{eqn:Eq5} 
\end{eqnarray}

\noindent
Its Fourier transform provides a bidirectional Poissonian distribution for the
electronic FCS $P_{el}(q)$ with corresponding left to right (right to left)
rates $\Gamma_{L \rightarrow R}$ ($\Gamma_{R \rightarrow L}$). Those coefficients are
evaluated within the scope of an approximation derived in Appendix B, which is
by construction \textit{current conserving} and \textit{consistent with the non-interacting
limit $g \rightarrow 0$}. We obtain in the case of symmetric coupling to the
leads $\Gamma_L = \Gamma_R = \Gamma/2$

\begin{eqnarray}  
\Gamma_{L \rightarrow R} &=& \frac{1}{4} \lbrace f_L
\Gamma_{Rh} + ( 1 - f_R )\Gamma_{Le} \rbrace(\tilde{\omega}_d)  \label{eqn:Eq6}   \\
\Gamma_{R \rightarrow L} &=& \frac{1}{4} \lbrace f_R
\Gamma_{Lh} + ( 1 - f_L )\Gamma_{Re} \rbrace (\tilde{\omega}_d)   \label{eqn:Eq7}  
\end{eqnarray}

\noindent
In Eq.\ref{eqn:Eq6} and \ref{eqn:Eq7}, the electronic and hole rates for
multi-phonon processes in the lead $X=L,R$ are respectively defined as
$\Gamma_{Xe}(\omega)=\Gamma e^{-g^2}\sum_{n=0}^{+\infty}\frac{g^{2n}}{n!}f_X(\omega+n\omega_0)$
and $\Gamma_{Xh}(\omega)=\Gamma e^{-g^2}\sum_{n=0}^{+\infty}\frac{g^{2n}}{n!}\lbrack 1 -
f_X(\omega-n\omega_0) \rbrack$. It is interesting to notice that the left to
right and right to left
rates of the Poissonian distribution exhibit a non analytical behaviour in the
e-ph coupling strength $g$ at low temperature, \textit{i.e.} a non perturbative calculation in $g$
is necessary to derive correct results in the tunnel limit $\Gamma \rightarrow
0$. Compared to the existing results in the literature
\cite{FC_Blockade_Giant_Fano_Factors,FCS_Low_Gamma,FCS_Low_Gamma_2},
the expression for the CGF derived in Eq.\ref{eqn:Eq5} does not contain
interaction-induced corrections to the Poissonian distribution.
Such corrections were shown to arise from an avalanche (electron bunching) mechanism for the 
dynamics of the transferred electrons \cite{FCS_Low_Gamma} which is not accounted for
in the lowest order expansion presented in Eq.\ref{eqn:Eq4}.
Such terms will emerge from a non perturbative evaluation of the CGF 
(``all order in $\Gamma$'' dressing of the dot Green function), which is out of the scope of the present article \footnote{
Such resummation of the most divergent diagrams in the linked cluster expansion might be difficult to achieve,
because Wick theorem is not valid when dealing with the Hamiltonian of Eq.\ref{eqn:Eq2}. Any approximation based on 
a Dyson-like equation is thus not strictly justified and the approximation made (compared to the unknown exact result)
appears to be difficult to control.}. 

\subsubsection{1. Electronic current}

\begin{figure}
   \begin{center}
        \includegraphics[width=0.95\columnwidth]{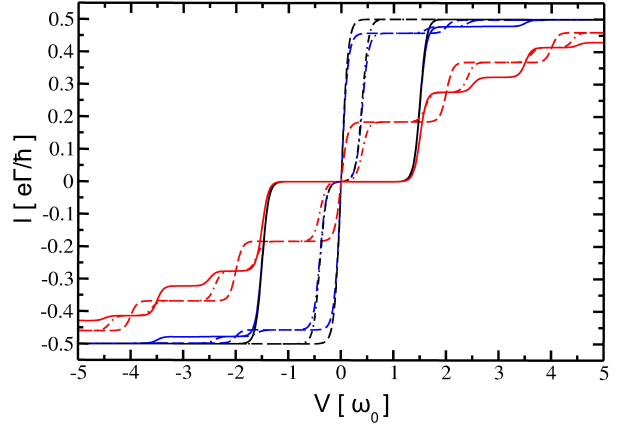}
        \caption{Current-voltage characteristics $I(V)$ in the scope of
          the approximation derived in Appendix B. The parameters used for
          this plot are $\Gamma=0.001 \omega_0$, $T=0.03 \omega_0$ and
          $\mu_L=-\mu_R=V/2$.  Black, blue and red curves correspond
          respectively to : $\lambda = 0.0; 0.3 ; 1.0 \omega_0$. Dashed,
          dashed-dotted and plain curves are obtained respectively with the dot position $\tilde{\omega}_d = 0.00; 0.20; 0.75 \omega_0$.}
        \label{fig:Current_Atomic_Limit}
   \end{center}
\end{figure}

The first cumulant $<q>^{(1)}=\Gamma_{L \rightarrow
  R}^{(I)}-\Gamma_{R\rightarrow L}^{(I)}$ corresponds to the mean current
$I(V)$ that flows across the junction. We show on
Fig.\ref{fig:Current_Atomic_Limit} typical $I(V)$ curves obtained by
varying the dot position and the e-ph coupling strength. For the case $\tilde{\omega}_d=0$, the $I(V)$ characteristics
exhibit an inelastic threshold at $V=\pm 2\omega_0$ corresponding to the
activation of inelastic tunnelling,
\textit{i.e.} an electron on the dot may tunnel to the leads by emitting a
phonon. Additional inelastic channels open when increasing $\lambda$ for voltages
multiple of $V=\pm n2\omega_0$, and correspond to the onset of multiple phonon
emission. It is interesting to notice that those inelastic patterns are
simply explained by Pauli principle that forbids multi-phonon transitions if
the final channel of diffusion is already occupied. In the
low-temperature limit $\Gamma \ll T \ll \omega_0$, the shape and magnitude
of the jumps at $V\approx\pm n2\omega_0$ is the result of Franck-Condon
factors \cite{Quantal_Transistor,Transport_I_V_molecules} entering into
Eq.\ref{eqn:Eq6} and \ref{eqn:Eq7}. Strictly speaking however,
our approximation breaks down in the limit $\tilde{\omega}_d \rightarrow 0$ (dashed curves of
Fig.\ref{fig:Current_Atomic_Limit}) corresponding to resonant
tunnelling. Only far from a resonant situation, namely when $|\tilde{\omega}_d|
\gg \Gamma$, is our low-$\Gamma$ (and thus low-transmission) approximation valid
(dashed-dotted and plain curves of
Fig.\ref{fig:Current_Atomic_Limit}). 
As explained in Appendix B.1, the transmission factor of the non interacting junction is approximated in
this limit by $T(\omega)\approx \pi \Gamma\delta(\omega-\tilde{\omega}_d)$, and the corresponding $I(V)$ curves
behave in the low-voltage region ($0 \le \tilde{\omega}_d,V \ll \omega_0$) as 

\begin{eqnarray}  
<q>^{(1)} \approx t_0 \frac{\Gamma e^{-g^2}}{2} \lbrace f_L(\tilde{\omega}_d) -
f_R(\tilde{\omega}_d) \rbrace \label{eqn:Eq8}  
\end{eqnarray}

\noindent
Eq.\ref{eqn:Eq8} is consistent with the Landauer-B\"{u}ttiker formula \cite{Landauer_Buttiker_Formula}, but with a
renormalised rate of tunnelling $\tilde{\Gamma} = \Gamma e^{-g^2}$ \footnote{For large
enough e-ph coupling strength $g$, the low-bias sequential conductance is exponentially
suppressed giving rise to a Franck-Condon blockade consistent with
Ref. \cite{FC_Blockade_Regime}. Higher-order terms in the $\Gamma$-expansion of
Eq.\ref{eqn:Eq4} should however not be neglected in order to derive
quantitative results and correspond to co-tunnelling
processes reported in Ref. \cite{Cotunneling_virtual_state,Inelastic_Cotunneling_CB_Device}.}. 
It fully coincides with it in the non-interacting case ($g \rightarrow 0$) and in the appropriate low-$\Gamma$ limit
stated above.

\subsubsection{2. Electronic current noise}

\begin{figure}
   \begin{center}
     \includegraphics[width=0.955\columnwidth]{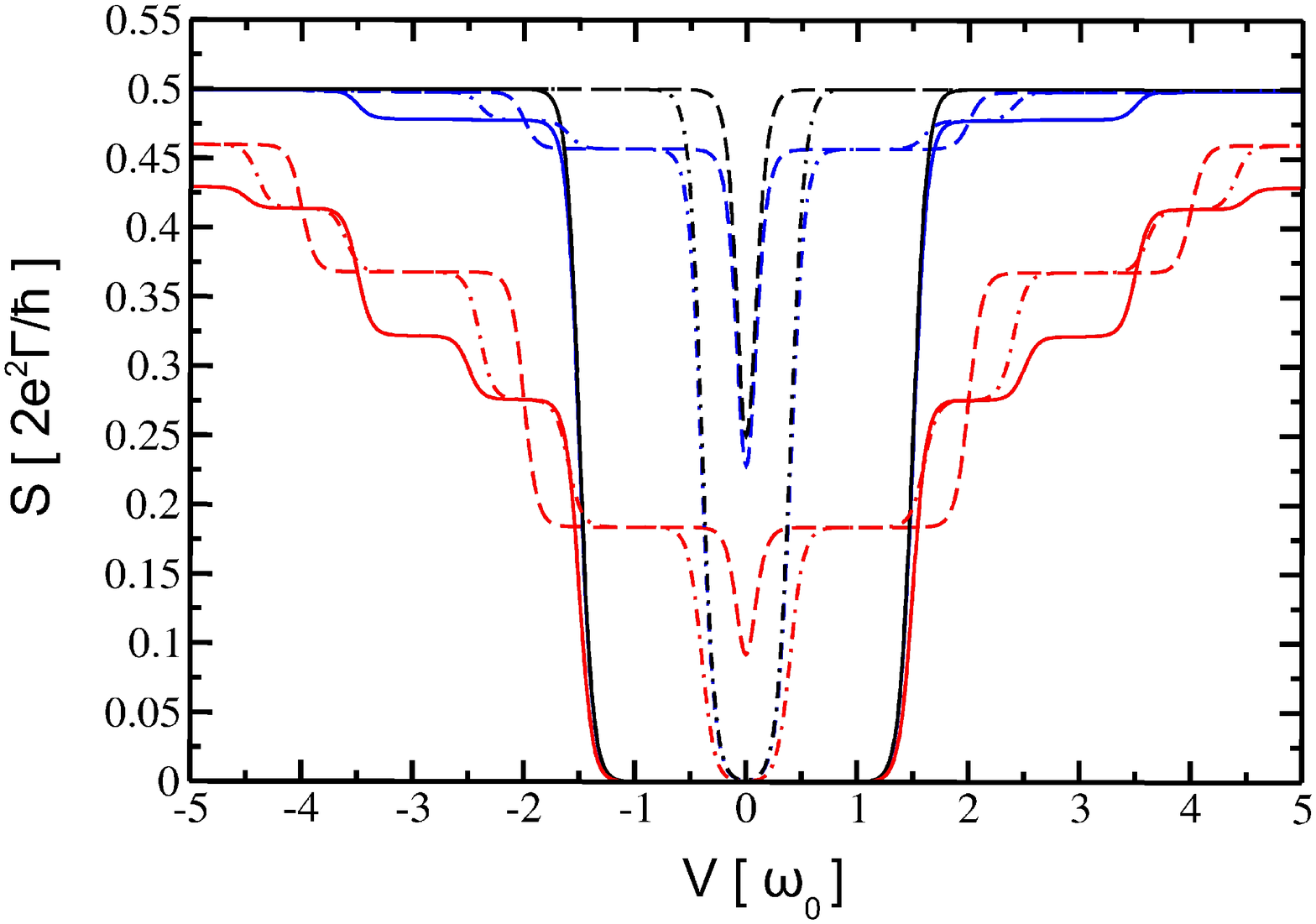}
        \caption{Noise-voltage characteristics $S(V)$ in the scope of
          the approximation derived in Appendix B. The parameters used for
          this plot are $\Gamma=0.001 \omega_0$, $T=0.03 \omega_0$ and
          $\mu_L=-\mu_R=V/2$. Black, blue and red curves correspond
          respectively to : $\lambda = 0.0; 0.3 ; 1.0 \omega_0$. Dashed,
          dashed-dotted and plain curves are obtained respectively with the dot position $\tilde{\omega}_d = 0.00; 0.20; 0.75 \omega_0$.}
        \label{fig:Noise_Atomic_Limit}
   \end{center}
\end{figure}

A similar behaviour (presence of inelastic thresholds) is observed for the case
of the second cumulant $\langle q^2 \rangle_c^{(1)} = \Gamma_{L \rightarrow
R}^{(I)}+\Gamma_{R\rightarrow L}^{(I)}$ that corresponds to the current-noise characteristics $S(V)$ on
Fig.\ref{fig:Noise_Atomic_Limit}. In the case of low-voltages ($0 \le
\tilde{\omega}_d,V \ll \omega_0$), the noise is given by 

\begin{eqnarray}  
<q^2>_c^{(1)} \approx t_0 \frac{\Gamma e^{-g^2}}{2} \lbrace f_L(1-f_R) +
f_R(1-f_L) \rbrace(\tilde{\omega}_d) \label{eqn:Eq9}  
\end{eqnarray}

\noindent
In the appropriate low-$\Gamma$ limit, Eq.\ref{eqn:Eq9} coincides with the scattering 
result for shot noise \cite{Shot_Noise_Mesoscopic_Conductors} when $g \rightarrow 0$.

\subsection{B. Population dynamics of the phonon mode}

\subsubsection{1. Derivation of the master equation for the phonons}

The computation of the phonon stationary distribution, namely the probability
$P_{ph}(n)$ of having $n$ phonons populating the mode in steady state, is more
difficult to achieve than
the corresponding calculation of the electronic FCS. The source of the
difficulty originates from the low-$\Gamma$ expansion of Eq.\ref{eqn:Eq4}
that does not take into account phonon emission and absorption on the same
footing, \textit{i.e.} Eq.\ref{eqn:Eq4} provides multi-phonon emission
processes at this order in the $\Gamma$-expansion but lacks higher-orders
absorption processes that are necessary to reach a steady state. \\
We adopt in the following a self-consistent
treatment that cures the problem by including in Eq.\ref{eqn:Eq4} any excited
state $\ket{m}$ of the local vibrational mode. More generally, we define 
$T_{ph}^{(m)}(n) = \sum_{q\in\mathbb{Z}} P^{(m)}(q,n)$ as the probability
of transition from the initial vibrational state $\ket{m}$ to the final state $\ket{m+n}$
during the measuring time $t_o$, after tracing out the
electronic degrees of freedom. This quantity is computed from the
generalised vibrational CGF, namely $\delta
S_{ph;m}^{(1)}(\xi)=\delta S^{(1)}_{m}(\chi=0,\xi)$, that includes the contribution
of vibronic Green functions $\mathcal{P}^{(m)}_\xi(t)= \langle T_c X(t)
X^{\dagger}(0)\rangle_m$ averaged other the excited phonon state ($m$ quanta
in the phonon mode). A detailed derivation of the generalised vibronic Green
functions and phonon CGF is proposed in Appendix A and C respectively. \\ 
In the absence of any external damping mechanism for the phonons (non equilibrated phonons),
the dynamics of the phonon population is self-determined for each time
interval $t_0$ by the electronic tunnelling mechanism as 

\begin{eqnarray} 
P_{ph}(t_0;n) = \sum_{m=0}^{+\infty} T_{ph}^{(m)}(n-m) P_{ph}(0;m)
\label{eqn:Eq10}  
\end{eqnarray}

\noindent 
In Eq.\ref{eqn:Eq10}, $P_{ph}(t_0;n)$ is the probability of having $n$
phonons populating the local vibrational mode at time $t_0$. The stationary distribution of the phonons
is obtained in the long time limit as the fixed point of Eq.\ref{eqn:Eq10}. In
the tunnel limit $\Gamma \rightarrow 0$, the transition rate $T_{ph}^{(m)}(n)$ is expanded in leading
order of $\Gamma$ as 

\begin{eqnarray} 
T_{ph}^{(m)}(n) &\approx& \delta_{n,0} + t_0 \Gamma_{ph}^{(m)}(n) + o(t_0^2,\Gamma^2) \label{eqn:Eq11} \\
\Gamma_{ph}^{(m)}(n) &=& -\int_{-\pi}^{\pi} \frac{d\xi}{2\pi} \frac{1}{t_0}
\delta S_{ph;m}^{(1)}(\xi) e^{-in\xi} \label{eqn:Eq12}
\end{eqnarray}

\noindent 
In Eq.\ref{eqn:Eq11}, $\Gamma_{ph}^{(m)}(n)$ is a transition rate per unit of
time corresponding to the multi-phonon process $\ket{m} \rightarrow
\ket{m+n}$. It is formally related to the Fourier
transform of the phonon CGF $S_{ph}^{(m)}(n)$ (see Eq.\ref{eqn:Eq12}). It is interesting to
notice that the matrix $[\Gamma_{ph}]_{n,m}=\Gamma_{ph}^{(m)}(n-m)$
might be evaluated analytically (see Appendix C) and has the property of
conserving the normalisation of the phonon distribution, \textit{i.e.} for
each index $m$ we have the following relation
amongst matrix elements $\sum_{n=0}^{+\infty} \Gamma_{ph;n,m} =
0$. Eq.\ref{eqn:Eq12} and \ref{eqn:Eq13} are an important result of this paper.
\textit{They are the constitutive relations that
connect the formulation of transport based on perturbation theory in Keldysh
space (usually used in the coherent transport regime) to the one based on the master equation in the tunnel limit.}
Including Eq.\ref{eqn:Eq11} into Eq.\ref{eqn:Eq10}, we derive from our formalism the standard quantum
master equation for the dynamics of the phonon population

\begin{widetext}
\begin{eqnarray} 
\dot{P}_{ph}(t;n) = \sum_{m=0;m \ne n}^{+\infty} \Gamma_{ph;n,m}
P_{ph}(t;m) - \sum_{m=0;m \ne n}^{+\infty} \Gamma_{ph;m,n} P_{ph}(t;n)
\label{eqn:Eq13}  
\end{eqnarray}
\end{widetext}

\subsubsection{2. Results for the phonon stationary distribution}

\begin{figure}[!ht]
        \includegraphics[width=0.96\columnwidth]{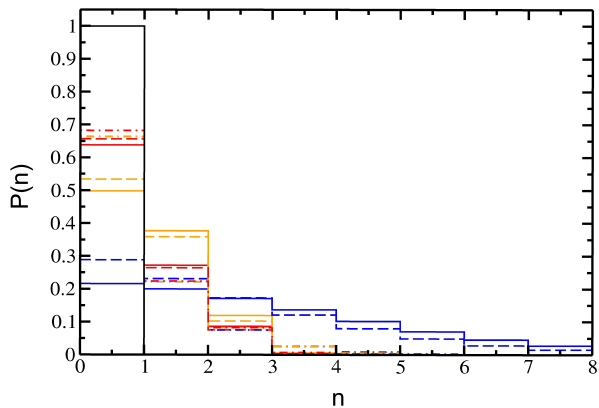}
        \caption{ Stationary distribution of the phonon population $P_{ph}(n)$ for a
          symmetrically biased molecular junction with parameters
          $\Gamma=0.001 \omega_0$, $T=0.03 \omega_0$ and
          $V=2.5 \omega_0$. Blue, orange and red curves are obtained for :
          $\lambda = 0.3; 1.0; 2.0 \omega_0$ respectively. Plain, dashed and
          dashed-dotted curves correspond respectively to the dot position
          $\tilde{\omega}_d = 0.00; 0.20; 0.75 \omega_0$. The black curve
          corresponds to $V=0.5 \omega_0$, $\tilde{\omega}_d = 0.00$ and
          $\lambda = 0.3 \omega_0$.}
        \label{fig:Phonon_Population}
\end{figure}

As an example, we compute the phonon stationary
distribution $P_{ph}(n)=\lim_{t\rightarrow +\infty} P_{ph}(t;n)$
obtained as the zero eigenvector of the $[\Gamma_{ph}]$
matrix (see Eq.\ref{eqn:Eq13}). For voltages below
the inelastic threshold ($V=0.5\omega_0$ for the black curve of
Fig.\ref{fig:Phonon_Population}), phonon emission is
forbidden by Pauli principle and the phonon mode is not populated
whatever the strength of e-ph coupling. In this case, the distribution $P_{ph}(n)$ is
a peak at $n=0$ given by the Bose equilibrium distribution.
For voltages above the inelastic threshold, a pumping mechanism appears : the onset of phonon
emission strongly drives the population of the vibrational mode out of equilibrium. For weak e-ph
coupling ($\lambda=0.3\omega_0$) and $V=2.5\omega_0$ (see blue curves on
Fig.\ref{fig:Phonon_Population}), the distribution $P_{ph}(n)$ is long
tailed, whereas for increasing $\lambda=1.0-2.0\omega_0$ (see orange-red curves on
Fig.\ref{fig:Phonon_Population}), it gets closer to the equilibrium
distribution (the tail is shorten). The corresponding non-monotonous behaviour of $P_{ph}(n)$ with $\lambda$ is
resulting from the competition between the emission-absorption
mechanism given by Eq.\ref{eqn:Eq13} and the selection rules imposed by the
transition matrix $[\Gamma_{ph}]$ (when $\lambda$ increases, desexcitation of
highly-excited states toward the ground
state are more likely to happen\cite{Quantal_Transistor}). 

\section{Conclusion-open questions}

In this work, we have developed a theoretical framework that enables to treat
on the same footing the concept of counting statistics of transmitted (fermionic)
charges commonly used in the field of quantum transport and the concept of counting
statistics of phonons (bosons).
We illustrated this framework by studying in detail the transport properties of an
interacting molecular junction, from both point of view of counting statistics
of transmitted electrons and excited phonons. We derived non-perturbative
results in the e-ph coupling strength which are valid in the lowest order of the
tunnelling rate to the leads $\Gamma$ (tunnel regime). Our description provides
a general framework  that unifies previous studies based on rate equations (in the
tunnel regime) to the one based on non-equilibrium Green functions techniques
(in the coherent regime).
\\

A natural extension and open question arising from the present work is to
investigate the role of higher-order terms in the $\Gamma$-expansion of
Eq.\ref{eqn:Eq4}. In particular, co-tunnelling processes
\cite{Cotunneling_virtual_state,Inelastic_Cotunneling_CB_Device} are expected
to be important in order to describe off-resonant transport into the deep
Franck-Condon blockade \cite{FC_Blockade_Regime} or to derive results valid in
the limit of large transmission (resonant tunnelling) \cite{Cotunneling_Resonance,Resonant_inelastic_tunneling_molecular_junctions}. Following the recent work of Maier \textit{et al} \cite{Charge_transfer_statistics_strong_eph_interaction}, such an
issue could be investigated by using a self-consistent scheme to compute the FCS in the strong coupling regime. 
More generally, the nature of transport properties and mechanisms in the polaron
crossover regime is an important, unsolved problem \cite{Polaron_Crossover} that deserves future
investigations following the lines drawn in the present article. 

\section{Acknowledgments}

R. Avriller is grateful to A. Levy Yeyati, D.F. Urban and A. Martin-Rodero for carefully
reading the manuscript and providing many interesting
and fruitful discussions. Financial support from the Spanish MICINN under contract
NAN2007-29366-E (CHENANOM) is acknowledged.

\section{Appendix}

\subsection{Appendix A : free Green functions}

\subsubsection{1. Electronic Green functions}

In this part, we compute the non-interacting electronic Green functions of the dot
and of the $X=L,R$ leads, namely $\hat{g}_{dd}(\omega)$
and $\hat{g}_{XX}(\omega)$. \\
We suppose that the leads are maintained in equilibrium and characterised by a flat
density of states $\rho_0$ and a chemical potential $\mu_{L(R)}=+(-)V/2$. The corresponding free Green functions
are given by

\begin{eqnarray} 
\hat{g}_{XX}(\omega) = \frac{i}{W} \left[ \begin{array}{cc}
2f_X-1 & 2f_X \\
-2(1-f_X) & 2f_X-1 \end{array} \right](\omega) \label{eqn:Eq14}  
 \end{eqnarray} 

\noindent
where $f_X(\omega)$ is the Fermi distribution of the lead $X=L,R$ and
$W=1/\pi\rho_0$ its bandwidth. \\
The free (non interacting) dot Green function is determined in the case of
symmetric contacts to the leads $\Gamma_L=\Gamma_R=\Gamma/2$ as

\begin{eqnarray} 
g_{dd}^{\alpha\alpha}(\omega) &=& \frac{\alpha (\omega - \omega_d) + i\Gamma
  (f_L+f_R-1)}{\Delta_{\chi}}(\omega) \label{eqn:Eq15}   \\ 
g_{dd}^{+-}(\omega) &=& i\Gamma \frac{e^{i\chi/2}f_L
  +e^{-i\chi/2}f_R}{\Delta_{\chi}}(\omega) \label{eqn:Eq16}   \\
g_{dd}^{-+}(\omega) &=& -i\Gamma \frac{e^{-i\chi/2}(1-f_L)
  +e^{i\chi/2}(1-f_R)}{\Delta_{\chi}}(\omega) \label{eqn:Eq17}  
\end{eqnarray} 

\noindent
where the Keldysh determinant is given by

\begin{eqnarray} 
&&\Delta_{\chi}(\omega)=(\omega-\omega_d)^2 + \Gamma^2 + \label{eqn:Eq18} \\
&&\Gamma^2 \lbrace (e^{i\chi}-1)f_L(1-f_R) + (e^{-i\chi}-1)f_R(1-f_L)
\rbrace(\omega) \nonumber 
\end{eqnarray} 

\noindent
The mean population of the dot $<n_d>$ is obtained after integration of the
non-diagonal component of the dot Green function $g_{dd}^{+-}(\omega)$ expressed 
at zero counting field   

\begin{eqnarray}
<n_d> = \int \frac{d\omega}{2\pi}
\frac{\Gamma}{(\omega-\omega_d)^2+\Gamma^2}\lbrace f_L(\omega) + f_R(\omega)
\rbrace \label{eqn:Eq19}  
\end{eqnarray}

\noindent
In the atomic limit ($\Gamma \rightarrow 0$), far from resonance ($|\omega_d| \gg
\Gamma$), the Lorentzian-integrand of Eq.\ref{eqn:Eq19} is approximated by a delta
function and we obtain for the voltage dependent mean population of the dot  

\begin{eqnarray}
\lim_{\Gamma \rightarrow 0} <n_d> = \frac{1}{2} \lbrace f_L(\omega_d)
+ f_R(\omega_d) \rbrace \label{eqn:Eq20}  
\end{eqnarray}

\noindent
Similarly, the atomic limit of the non-interacting Green function is obtained as

\begin{eqnarray} 
g_{dd}^{\alpha\alpha}(\omega) &=& \alpha \lbrace
\frac{1-<n_d>}{\omega-\omega_d+\alpha i\eta}+
\frac{<n_d>}{\omega-\omega_d-\alpha i\eta} \rbrace \label{eqn:Eq21}   \\ 
g_{dd}^{+-}(\omega) &=& i2\pi<n_d> \delta(\omega-\omega_d) \label{eqn:Eq22}   \\
g_{dd}^{-+}(\omega) &=& -i2\pi(1-<n_d>) \delta(\omega-\omega_d) \label{eqn:Eq23}  
\end{eqnarray} 

\subsubsection{2. Vibronic Green functions}

The free vibronic Green function $\mathcal{P}_{\xi=0}(t)= \langle T_c X(t)
X^{\dagger}(0)\rangle_0$ is evaluated by performing the
average other the phonon ground state ($m=0$ quanta in the vibrational
mode). Using the Glauber equality \cite{Photon_Correlations} $e^{C+D} = e^C e^D e^{-\frac{1}{2}\lbrack
  C,D \rbrack}$ which is valid for any operators $(C,D)$ that commute with
their commutator, we obtain similarly to Ref. \cite{Interpolative_self_energies} 

\begin{eqnarray}
\mathcal{P}_{\xi=0}^{\alpha\alpha}(\omega) &=& e^{-g^2} \sum_{n=0}^{+\infty}
\frac{g^{2n}}{n!} \lbrace \frac{1}{\omega-\alpha n\omega_0 +i\eta} - \label{eqn:Eq24} \\
&&\frac{1}{\omega+\alpha n\omega_0 -i\eta} \rbrace \nonumber \\
\mathcal{P}_{\xi=0}^{\alpha-\alpha}(\omega) &=& e^{-g^2} \sum_{n=0}^{+\infty}
\frac{g^{2n}}{n!} 2\pi \delta(\omega+\alpha n\omega_0) \label{eqn:Eq25}
\end{eqnarray}
 
\noindent
More generally, we define the $\xi$-dependent free vibronic Green function
as $\mathcal{P}^{(m)}_\xi(t)= \langle T_c X(t) X^{\dagger}(0)\rangle_m$, where
the mean value is performed other an excited phonon state ($m \ne
0$ quanta in the mode). We obtain in general 

\begin{eqnarray}
\mathcal{P}_\xi^{(m);\alpha\alpha}(t) &=&
\theta(t)C^{\alpha\alpha}_{\alpha;m}(t)+\theta(-t)C^{\alpha\alpha}_{-\alpha;m}(t)
 \label{eqn:Eq26} \\
\mathcal{P}_\xi^{(m);\alpha-\alpha}(t) &=& C^{-+}_{-\alpha;m}(t) \label{eqn:Eq29} 
\end{eqnarray}

\noindent
where the correlators $C^{\alpha\beta}_{1;m}(t)=\bra{m} X_{\alpha\xi}(t)
X_{\beta\xi}^{\dagger}(0) \ket{m}$ and $C^{\alpha\beta}_{-1;m}(t)=\bra{m} X_{\alpha\xi}^{\dagger}(0)
X_{\beta\xi}(t) \ket{m}$ are given in Fourier representation by  

\begin{widetext}
\begin{eqnarray}
C^{\alpha\beta}_{\pm 1;m}(\omega) = 2\pi e^{-g^2}
\sum_{p_1=0}^{+\infty}\sum_{p_2=0}^{m}\sum_{p_3=0}^{2p_2}
\frac{g^{2(p_1+p_2)}}{p_1!p_2!} C^{m}_{p_2} C^{2p_2}_{p_3}(-1)^{p_3}
e^{-i\frac{\alpha-\beta}{2}(p_1+p_2-p_3)\xi} 
\delta \lbrack \omega \mp (p_1+p_2-p_3)\omega_0 \rbrack \label{eqn:Eq30}
\end{eqnarray}
\end{widetext}

\noindent
It is interesting to notice that only non-diagonal components of the phonon
propagator are explicitly dependent on the phonon counting field. If the
phonon states are restricted to the $m=0$ ground state (like in the low-$\Gamma$ result of Eq.\ref{eqn:Eq4}),
the only possible processes available correspond
to multiple phonon emission. Only when considering an excited phonon state ($m \ge 1$) are
absorption processes allowed.

\subsection{Appendix B : rates of the Poisson distribution}

\subsubsection{1. The non interacting case}

Using our formalism and the expression of the dot Green functions (see
Eq.\ref{eqn:Eq15}-\ref{eqn:Eq17}), the electronic CGF of the resonant level in the non-interacting case ($g=0$)
might be determined in all orders of $\Gamma$. This provides the Levitov-Lesovik result \cite{Hamiltonian_FCS_Levitov_Formula,Hamiltonian_FCS_Levitov_Formula_2,FCS_Gogolin} for the free electronic CGF $S^{(0)}_{el}(\chi)$, and 
the associated binomial FCS of transmitted electrons 

\begin{eqnarray}
&&S^{(0)}_{el}(\chi) = -t_0 \int \frac{d\omega}{2\pi} \ln \Big{\lbrace} 1 + T(\omega)\Big{\lbrack} \label{eqn:Eq31} \\
&& (e^{i\chi}-1)f_L(1-f_R) + (e^{-i\chi}-1)f_R(1-f_L) \Big{\rbrack}(\omega) \Big{\rbrace}\nonumber
\end{eqnarray}

\noindent
Eq.\ref{eqn:Eq31} is associated with resonant tunnelling of electrons and holes
through the dot, characterised by a transmission factor

\begin{eqnarray}
T(\omega)=\frac{\Gamma^{2}}{(\omega-\omega_d)^2+\Gamma^2} \label{eqn:Eq32} 
\end{eqnarray}

\noindent
The limit of low-$\Gamma$ and off-resonant situation $|\omega_d| \gg \Gamma$
corresponds to an approximate transmission factor $T(\omega)\approx \pi
\Gamma\delta(\omega-\omega_d)$. The corresponding expansion of the electronic
CGF in the lowest order of $\Gamma$ provides the Poissonian result $S^{(0)}_{el}(\chi) \approx -t_0 \lbrace \Gamma_{L \rightarrow R}^{(0)}(e^{i\chi}-1) +
\Gamma_{R \rightarrow L}^{(0)}(e^{-i\chi}-1) \rbrace$ with rates expressed as

\begin{eqnarray}
\Gamma_{L \rightarrow R}^{(0)} &=& \frac{\Gamma}{2} f_L(\omega_d)\lbrack
1-f_R(\omega_d) \rbrack \label{eqn:Eq33} \\
\Gamma_{R \rightarrow L}^{(0)} &=& \frac{\Gamma}{2} f_R(\omega_d)\lbrack
1-f_L(\omega_d) \rbrack \label{eqn:Eq34}
\end{eqnarray}

\noindent
In the following, the results for the interacting case have to be understood as well in the
appropriate low-$\Gamma$ and low-transmission limit stated above. 

\subsubsection{2. Interacting case : choice of the gauge}

When evaluating the lowest order in the $\Gamma$-expansion of the electronic
CGF, one is not ensured (if no self-consistency is achieved) that current
conservation is fulfilled, namely that
the obtained cumulants versus voltage curves have a well defined symmetry
under the transformation $V \rightarrow -V$, nor that the obtained CGF
coincides with the non-interacting result of Eq.\ref{eqn:Eq33} and
Eq.\ref{eqn:Eq34} when $g \rightarrow 0$. \\
We use this constraint of both \textit{charge conservation} and \textit{recovering
the non-interacting Levitov-Lesovik result} to select the proper gauge in
implementing the electronic counting field in the Hamiltonian. Clearly, the
choice of a gauge that incorporates the electronic counting field in the left electrode only $t_{Ld} \rightarrow t_{Ld}
e^{-i\chi(t)/2}$ breaks the symmetry between left and right electrode when
expanding the CGF to the lowest order in $\Gamma$, \textit{i.e.} this choice
of gauge is not current conserving. The symmetry between both electrodes (and
hence current conservation) is restored by introducing the electronic counting
field symmetrically in both leads, namely $t_{L(R)d} \rightarrow t_{L(R)d}
e^{\mp i\chi(t)/4}$. In the following, we make this choice of a symmetric gauge
that is by construction current conserving. \\
We then evaluate the electronic CGF from Eq.\ref{eqn:Eq4}, using
Eq.\ref{eqn:Eq24}-\ref{eqn:Eq25} for the free phonon propagator  

\begin{widetext}
\begin{eqnarray}
\delta S^{(1)}_{el}(\chi) \approx -\frac{t_0}{2} W e^{-g^2} \sum_{n=0}^{+\infty}
\frac{g^{2n}}{n!} \sum_{\alpha=\pm} \int \frac{d\omega_1}{2\pi} \Big{\lbrace}
g^{\alpha-\alpha}_{dd}(\omega_1) \Big{\lbrack} \Gamma_L e^{-i\alpha\chi/2} 
g^{-\alpha\alpha}_{LL} + \Gamma_R e^{i\alpha\chi/2} g^{-\alpha\alpha}_{RR}
\Big{\rbrack}(\omega_1-\alpha n\omega_0) \Big{\rbrace} \label{eqn:Eq35}
\end{eqnarray}
\end{widetext}

\noindent 
In Eq.\ref{eqn:Eq35}, the free dot Green function $\hat{g}_{dd}(\omega)$
has to be evaluated. However, by implementing na\"{i}vely the bare dot Green
function given in the atomic limit by Eq.\ref{eqn:Eq21}-\ref{eqn:Eq23}, the obtained CGF
although current conserving does not reproduce the limiting case of the
Levitov-Lesovik formula when $g \rightarrow 0$. This is related to the fact that the non equilibrium 
state of the dot level subsystem is ill defined in the atomic limit $\Gamma \rightarrow 0$, \textit{i.e.}
one has to artificially include the presence of electrodes maintained under a constant voltage bias as 
an external boundary term (for instance, as a voltage dependent population of the dot in the expression of the 
bare dot Green function in Eq.\ref{eqn:Eq21}-\ref{eqn:Eq23}). We have found the following
procedure to overcome this difficulty and compute safely the electronic CGF in such a way that the non-interacting
limit for the rates as given by Eq.\ref{eqn:Eq33}-\ref{eqn:Eq34} is recovered. We
first use the expression of the free dot Green function in all orders of
$\Gamma$ as written in Eq.\ref{eqn:Eq15}-\ref{eqn:Eq17} and obtain for the
CGF 

\begin{widetext}
\begin{eqnarray}
\delta S^{(1)}_{el}(\chi) &=& -\frac{t_0}{2} \int \frac{d\omega \Gamma^2}{2\pi
  \Delta_{\chi}} \Big{\lbrace} f_LA_{Lh} + (1-f_L)A_{Le} + f_RA_{Rh} +
(1-f_R)A_{Re} +  \label{eqn:Eq36} \\
&&e^{i\chi} \Big{\lbrack} f_LA_{Rh}+ (1-f_R)A_{Le} \Big{\rbrack}
+ e^{-i\chi} \Big{\lbrack} f_RA_{Lh} + (1-f_L)A_{Re} \Big{\rbrack} 
\Big{\rbrace}(\omega) \nonumber 
\end{eqnarray}
\end{widetext}

\noindent  
We finally evaluate the Keldysh determinant $\Delta_{\chi}(\omega)$ at zero
electronic counting field $\chi=0$ and go to the limit $\Gamma \rightarrow 0$.
We obtain Eq.\ref{eqn:Eq5}-\ref{eqn:Eq7} for the electronic CGF and its corresponding rates.

\subsection{Appendix C : transition rates for the phonons}

In this Appendix, we derive analytical expressions for the generalised phonon
CGF  $\delta S_{ph;m}^{(1)}(\xi)$  and for the corresponding multi-phonon transition rates per unit of
time $\Gamma_{ph}^{(m)}(n)$. We first write Eq.\ref{eqn:Eq4} in terms of the
generalised phonon Green functions defined in Appendix A 

\begin{widetext}
\begin{eqnarray}
\delta S^{(1)}_{ph;m}(\xi) &\approx& -\frac{t_0}{2} W e^{-g^2}
\sum_{p_1=0}^{+\infty}\sum_{p_2=0}^{m}\sum_{p_3=0}^{2p_2}
\frac{g^{2(p_1+p_2)}}{p_1!p_2!} C^{m}_{p_2} C^{2p_2}_{p_3}(-1)^{p_3}
\Big{\lbrack} e^{i(p_1+p_2-p_3)\xi} - 1 \Big{\rbrack}  \label{eqn:Eq37} \\
&&\sum_{\alpha=\pm} \int \frac{d\omega_1}{2\pi} \Big{\lbrace}
g^{\alpha-\alpha}_{dd}(\omega_1) \Big{\lbrack} \Gamma_L 
g^{-\alpha\alpha}_{LL} + \Gamma_R g^{-\alpha\alpha}_{RR}
\Big{\rbrack}(\omega_1-\alpha (p_1+p_2-p_3)\omega_0) \Big{\rbrace} \nonumber
\end{eqnarray}
\end{widetext}

\noindent
Putting the evaluation of the leads and of the dot Green functions
as given by Eq.\ref{eqn:Eq14} and
Eq.\ref{eqn:Eq21}-\ref{eqn:Eq23}, we obtain after Fourier
transforming Eq.\ref{eqn:Eq37} the following expression for the transition
rates

\begin{widetext}
\begin{eqnarray}
\Gamma_{ph}^{(m)}(n) &=& e^{-g^2} \sum_{p_1=0}^{+\infty}\sum_{p_2=0}^{m}\sum_{p_3=0}^{2p_2}
\frac{g^{2(p_1+p_2)}}{p_1!p_2!} C^{m}_{p_2} C^{2p_2}_{p_3}(-1)^{p_3} \Big{\lbrace} \nonumber \\
&&<n_d> \lbrack \Gamma_L(1-f_L) + \Gamma_R(1-f_R)\rbrack (\omega_{-}) + \nonumber \\
&&(1-<n_d>) \lbrack \Gamma_Lf_L + \Gamma_Rf_R \rbrack (\omega_{+}) \Big{\rbrace} \Big{\lbrack}
\delta_{n,p_1+p_2-p_3} -\delta_{n,0} \Big{\rbrack} \label{eqn:Eq38}
\end{eqnarray}
\end{widetext}

\noindent
where $\omega_{\pm} = \tilde{\omega}_d \pm (p_1+p_2-p_3)\omega_0$.
The constraints on summations due to the last term in Eq.\ref{eqn:Eq38} imply
that the transition rates are conserving the phonon probability, namely that 
$\sum_{n=0}^{+\infty} \Gamma_{ph;n,m} = 0$.



\begin{thebibliography}{}

\bibitem{Hamiltonian_FCS_Levitov_Formula}
L. S. Levitov, H.-W. Lee, and G. B. Lesovik,
J. Math. Phys. \textbf{37}, 4845 (1996).

\bibitem{Hamiltonian_FCS_Levitov_Formula_2}
L. S. Levitov and M. Reznikov,  
Phys. Rev. B \textbf{70}, 115305 (2004).

\bibitem{Photon_Correlations}
R. J. Glauber,
Phys. Rev. Lett. \textbf{10}, 84 (1963).

\bibitem {Quantum_Noise_Mesoscopic_Physics}
\emph{Quantum Noise in Mesoscopic Physics}, 
edited by Y.V. Nazarov, NATO Science Series, II. Mathematics, Physics 
and Chemistry, vol \textbf{97}, ISBN 1-4020-1239-X (2002).

\bibitem{Electron_Transport_Molecular_Wire_Junctions}
A. Nitzan and M.A. Ratner,
Science \textbf{300}, 1384-1389 (2003).

\bibitem{Shot_Noise_Mesoscopic_Conductors} 
Y.M. Blanter, M. Buttiker,
Phys. Rep. \textbf{336}, 1 (2000).

\bibitem{Quantum_Measurement_Phonon_Shot_Noise}
A.A. Clerk, F. Marquardt and J.G.E. Harris,
Phys. Rev. Lett. \textbf{104}, 213603 (2010).

\bibitem{Charge_Tunneling_Nanojunctions}
\emph{Single Charge Tunneling}, edited by G.-L. Ingold and Y.V. Nazarov,
NATO ASI Series B, Vol. \textbf{294}, pp. 21-107 (Plenum Press, New York, 1992).

\bibitem{FC_Blockade_Regime}
J. Koch, F. von Oppen and A.V. Andreev,
Phys. Rev. B  \textbf{74}, 205438 (2006). 

\bibitem{FC_Blocakde_Suspended_CNTs}
R. Leturcq, C. Stampfer, K. Inderbitzin, L. Durrer, C. Hierold, E. Mariani, 
M.G. Schultz, F. von Oppen and  K. Ensslin,
Nature Physics \textbf{5}, 327-331 (2009). 

\bibitem{Transport_I_V_molecules}
J. Koch and F. von Oppen,
Phys. Rev. B \textbf{72}, 113308 (2005). 

\bibitem{FC_Blockade_Giant_Fano_Factors}
J. Koch and F. von Oppen,
Phys. Rev. Lett. \textbf{94}, 206804 (2005).

\bibitem{FCS_Low_Gamma}
J. Koch, M.E. Raikh and F. von Oppen,
Phys. Rev. Lett. \textbf{95}, 056801 (2005).

\bibitem{FCS_Low_Gamma_2}
B. Dong, H.Y. Fan, X.L. Lei and N.J.M. Horing,
J. Appl. Phys. \textbf{105}, 113702 (2009).    

\bibitem{FCS_charge_shuttle}
F. Pistolesi,
Phys. Rev. B \textbf{69}, 245409 (2004).

\bibitem{Atomic_Chains}
N. Agrait, C. Untiedt, G. Rubio-Bollinger, and S. Vieira,
Phys. Rev. Lett. \textbf{88}, 216803 (2002).

\bibitem{Molecular_junctions}
O. Tal, M. Krieger, B. Leerink, and J.M. van Ruitenbeek,
Phys. Rev. Lett. \textbf{100}, 196804 (2008).

\bibitem{Keldysh_Formalism}
L.V. Keldysh, Zh. Eksp. Teor. Fiz. \textbf{47}, 1515 (1964),
[Sov. Phys. JETP \textbf{20}, 1018 (1965)].

\bibitem{Quantal_Transistor}
A. Mitra, I. Aleiner, and A. J. Millis,
Phys. Rev. B \textbf{69}, 245302 (2004).

\bibitem{Inelastic_electron_tunneling_spectroscopy}
M. Galperin. M. A. Ratner and A. Nitzan,
J. Chem. Phys. \textbf{121}, 11965 (2004).

\bibitem{Voltage_induced_singularities_transport}
O. Entin-Wohlman, Y. Imry, and A. Aharony,
Phys. Rev. B \textbf{80}, 035417 (2009).

\bibitem{Inelastic_Scattering_Local_Heating_Atomic_Gold_Wires}
T. Frederiksen, M. Brandbyge, N. Lorente, and A.-P. Jauho,
Phys. Rev. Lett. \textbf{93}, 256601 (2004).

\bibitem{Electron-vibration_interaction_transport_atomic_gold_wires}
J. K. Viljas, J. C. Cuevas, F. Pauly, and M. Hafner,
Phys. Rev. B \textbf{72}, 245415 (2005).

\bibitem{Universal_features_electron_phonon_interactions_atomic_wires}
L. de la Vega, A. Martin-Rodero, N. Agrait, and A. Levy Yeyati,
Phys. Rev. B. \textbf{73}, 075428 (2006).

\bibitem{Theory_current_shot-noise}
J.-X. Zhu and A.V. Balatsky,
Phys. Rev. B. \textbf{67}, 165326 (2003).

\bibitem{Inelastic_tunneling_effects_noise}
M. Galperin, A. Nitzan, and M.A. Ratner,
Phys. Rev. B. \textbf{74}, 075326 (2006).

\bibitem{FCS_Molecular_junctions}
T. L. Schmidt and A. Komnik,
Phys. Rev. B. \textbf{80}, 041307(R) (2009).

\bibitem{FCS_Molecular_junctions_2}
R. Avriller and A. Levy Yeyati,
Phys. Rev. B. \textbf{80}, 041309(R) (2009).

\bibitem{FCS_Molecular_junctions_3}
F. Haupt, T. Novotny, and W. Belzig,
Phys. Rev. Lett. \textbf{103}, 136601 (2009).

\bibitem{FCS_Molecular_junctions_Role_Phonon_Fluctuations}
D. F. Urban, R. Avriller and A. Levy Yeyati,
Phys. Rev. B \textbf{82}, 121414(R) (2010). 

\bibitem{Lang_Firsov_Transformation}
I. G. Lang and Yu. A. Firsov,
Zh. Eksp. Teor. Fiz. \textbf{43}, 1843 (1962).
[Sov. Phys. JETP \textbf{16}, 1301 (1963)].

\bibitem{FCS_Gogolin}
A.O. Gogolin and A. Komnik,
Phys. Rev. B. \textbf{73}, 195301 (2006).

\bibitem{Landauer_Buttiker_Formula}
M. B\"{u}ttiker, Y. Imry, R. Landauer and S. Pinhas, 
Phys. Rev. B \textbf{31}, 6207-6215 (1985).

\bibitem{Cotunneling_virtual_state}
D.V. Averin and Yu. V. Nazarov,    
Phys. Rev. Lett. \textbf{65}, 2446-2449 (1990).

\bibitem{Inelastic_Cotunneling_CB_Device}
K. Flensberg,  
Phys. Rev. B \textbf{55}, 13118-13123 (1997).

\bibitem{Cotunneling_Resonance}
J. K\"{o}nig, H. Schoeller and G. Sch\"{o}n,  
Phys. Rev. Lett. \textbf{78}, 4482-4485 (1997).

\bibitem{Resonant_inelastic_tunneling_molecular_junctions}
M. Galperin, A. Nitzan, and M.A. Ratner,
Phys. Rev. B \textbf{73}, 045314 (2006).

\bibitem{Charge_transfer_statistics_strong_eph_interaction}
S. Maier, T.L. Schmidt, A. Komnik,
Phys. Rev. B. \textbf{83}, 085401 (2011).

\bibitem{Polaron_Crossover}
A. Zazunov and T. Martin,  
Phys. Rev. B \textbf{76}, 033417 (2007). 

\bibitem{Interpolative_self_energies}
A. Martin-Rodero, A. Levy Yeyati, F. Flores, and R. C. Monreal,
Phys. Rev. B \textbf{78}, 235112 (2008). 

\end{thebibliography}
\end{document}